\documentclass[floatfix,superscriptaddress,aps,preprint,showpacs,amsmath,amssymb]{revtex4}
\usepackage{epsfig}
\usepackage{slashed}
\begin{document}
\title{On the Chern-Simons terms in Lifshitz like quantum electrodynamics}

\author{Van S\'ergio Alves}
\email{vansergi@ufpa.br}
\affiliation{Faculdade de F\'\i sica, Universidade Federal do Par\'a, 66075-110, Bel\'em, Par\'a, Brazil}

\author{B. Charneski}
\email{bruno@fma.if.usp.br}
\affiliation{Instituto de F\'{\i}sica, Universidade de S\~ao Paulo\\
Caixa Postal 66318, 05315-970, S\~ao Paulo, SP, Brazil}

\author{M. Gomes}
\email{mgomes@fma.if.usp.br}
\affiliation{Instituto de F\'{\i}sica, Universidade de S\~ao Paulo\\
Caixa Postal 66318, 05315-970, S\~ao Paulo, SP, Brazil}

\author{Leonardo Nascimento}
\email{lnascimento@ufpa.br}
\affiliation{Faculdade de F\'\i sica, Universidade Federal do Par\'a, 66075-110, Bel\'em, Par\'a, Brazil}

\author{Francisco Pe\~na}
\email{fcampos@ufro.cl}
\affiliation{Departamento de Ciencias F\'\i sicas, Facultad de Ingenier\'\i a, Ciencias y {
Administraci\'on}, Universidad de La Frontera, Avda. Francisco Salazar 01145, Casilla 54-D, Temuco, Chile}

\date{\today}

\begin{abstract}
In this work the generation of generalized Chern-Simons terms in three dimensional quantum electrodynamics with high spatial derivatives is studied.  We analyze the self-energy corrections to the gauge field propagator by considering an expansion of the corresponding amplitudes  up to third order in the external momenta. The divergences of the corrections are determined and explicit forms for the  Chern-Simons terms with high derivatives are obtained.
Some unusual aspects of the calculation are stressed and the existence of a smooth isotropic limit is proved.
 The transversality  of the anisotropic gauge propagator is also discussed. 
\end{abstract}


\maketitle

\section{Introduction}

A great deal of attention has been devoted to the analysis of  possible effects of the spacetime anisotropy \cite{cosmgravrefs,stringrefs,condmatrefs}. In the context of quantum field theory (QFT), this possibility appears as a tool to study nonrenormalizable theories since it improves the ultraviolet behavior of the perturbative series, in spite of violating the Lorentz symmetry which in its turn has been considered in several situations  \cite{ncrefs,Colladay:1998fq,Charneski:2008hy,LVtable}.

The different behavior  between  space and time coordinates $x^i\rightarrow bx^i$, $t\rightarrow b^zt$~\cite{anisorefs}, ameliorates the ultraviolet behavior and it has been argued  that  four-dimensional gravity becomes renormalizable when $z=3$ \cite{Horava:2009uw}.
However, the implementation of this kind of  anisotropy introduces unusual aspects and therefore it is important to carefully investigate  the consequences of this new approach \cite{mg2,Farias:2011aa}.

One special situation concerns the Chern-Simons (CS) term \cite{DeserJakiw,jackiw}:  when it is  added to  quantum electrodynamics (QED) Lagrangian, the generation of mass for the gauge field happens without gauge symmetry breaking and it  naturally emerges  from quantum corrections to the gauge field propagator \cite{redlich}. Beyond that, applications have been devised into diverse areas \cite{Charneski:2008hy,condmatCSrefs,hall,supercond,stringsCSrefs}. Therefore, the study of the spacetime anisotropy on the theories involving the CS term is certainly relevant.

In this work we will  analyze the new contributions to the self-energy of the gauge field in the  $z=2$ case up to one loop order in  the small momenta regime. In particular,  corrections to the CS term will be studied.  On general grounds we expect that the leading CS corrections have following form:
\begin{eqnarray}\label{CSexpec}
{\cal L_{CS}}&=&a\epsilon^{\mu\rho\nu}A_\mu\partial_\rho A_\nu+b\epsilon^{\mu\rho\nu}\Delta A_\mu\partial_\rho A_\nu+c\;\epsilon^{\mu\rho\nu}\partial_0^2 A_\mu\partial_\rho A_\nu,
\end{eqnarray}
where $\Delta$ denotes the Laplacian. In an interesting work \cite{Deser:1999pa} where the isotropic spacetime was considered, a CS term with structure similar to (\ref{CSexpec})  was employed. There,  the Laplacian was replaced by a D'Alambertian and $c = 0$, so that the gauge propagator shows two massive excitations, one of them being a ghost like.

By starting from (\ref{CSexpec}) with $c=0$, so that there is at most one time derivative, and adding the Maxwell term,
\begin{eqnarray}\label{isoM}
{\cal L_A}&=&-\frac{1}{4}F_{\mu\nu} F^{\mu\nu}-\frac{\lambda}{2}(\partial_\mu A^\mu)^2,
\end{eqnarray}
we obtain the propagator
\begin{eqnarray}\label{isoGP}
{D^{\mu\nu}}(k)&=&\frac{-i}{[k^2-(a-b\vec{k}^2)^2]}\left(g^{\mu\nu}-\frac{k^\mu k^\nu}{k^2}-\frac{i(a-b\vec{k}^2)\epsilon^{\mu\rho\nu}k_\rho}{k^2}\right)+\nonumber\\
&-&\frac{ik^\mu k^\nu}{\lambda k^2[k^2-(a-b\vec{k}^2)^2]} + \frac{i(a-b\vec{k}^2)^2 k^\mu k^\nu}{\lambda k^2[k^2-(a-b\vec{k}^2)^2]},
\end{eqnarray}
which does not contain particle like poles and, for small momenta, indicates disturbances propagating with squared velocity $(1-2ab)$. In this work we will also check the consistency with respect to the gauge symmetry of these corrections  through the verification of the transversality of the self-energy contributions to the gauge field propagator.
Our calculations show that, analogously to the relativistic situation, although finite the coefficient $a$ is regularization dependent. Actually, if dimensional reduction (see appendix \ref{appendixB}) is employed, the constant $a$ turns out to be equal to zero whereas in the relativistic situation it is nonvanishing.

This work is organized as follows. In Sec. \ref{anisotropic} we introduce  high derivative terms in the Dirac and Maxwell Lagrangians and analyze the CS generation and the characteristics of the self-energy  of the gauge field. For simplicity, our calculations  will be restricted to small momenta regime. In Sec. \ref{current} we discuss the transversality of the corrections to the gauge field propagator. Sec. \ref{remarks} presents some concluding remarks. Two appendices are dedicated to  detail some aspects of the calculations.


\section{Generation of Chern-Simons terms in the anisotropic QED}\label{anisotropic}

The modified Dirac Lagrangian, containing a high spatial derivative of second order is
\begin{eqnarray}\label{LDA}
{\cal L}=\bar\psi (i \gamma^0D_0)\psi+b_1\bar\psi (i \gamma^iD_i)\psi+b_2\bar\psi (i \gamma^iD_i)^2\psi-m\bar\psi\psi,
\end{eqnarray}
where $i=1,2$ and $D^\mu= \partial^\mu-ieA^\mu$ $(\mu=0,1,2)$ is the covariant derivative and $\gamma^{\mu}$ indicates a $2\times 2$ representation of the Dirac gamma matrices. To obtain the QED Lagrangian we add to (\ref{LDA}) the Maxwell term with high derivatives:
\begin{eqnarray}\label{lag-anisoM}
{\cal L_M}=\frac{1}{4}\left(F_{ij} \Delta F_{ij}+2F_{0i}F_{0i}\right)=\frac{1}{4}F_{ij}\Delta F_{ij}+\frac{1}{2}\partial_iA_0\partial_iA_0-\partial_0A_i\partial_iA_0+\frac{1}{2}\partial_0A_i\partial_0A_i,
\end{eqnarray}
in which $F_{ij}=\partial_i A_j-\partial_j A_i$.  The above contribution exhibits a mixture among space and time components, which may cause complications on the calculations involving the gauge field propagator. It is possible to avoid the mixed propagators by conveniently choosing the gauge fixing~\cite{Farias:2011aa}. As the gauge field propagator does not appear in our calculations we will keep  an inespecific gauge fixing, ${\cal L_{F}}$.  Despite of the mixing terms or gauge choice, notice that the determination of the transversal and longitudinal parts of the gauge field propagator is highly non trivial in anisotropic theory,  thus we will dedicate a section to an analyses of the transversality of the corrections to the gauge field propagator.

\begin{figure}[h]
\begin{center}
\includegraphics[height=3.5cm]{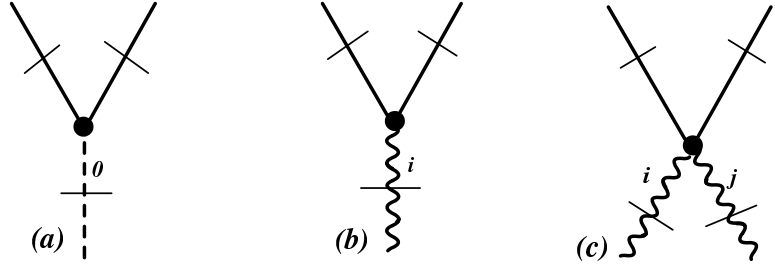}
\caption{Graphical representation of the interaction vertices. The continuos line stands for the fermion field, the dashed and wavy lines are for the $A_0$ and $A_i$ components, respectively. The diagram (a) is the trilinear vertex with two fermion fields and the $A_0$ component.  Figure (b) is the vertex  with two fermion fields and the $A_i$ component; besides that, it may contain a spatial derivative. Graph (c) is the quadrilinear vertex with two fermion fields and two $A_i$ components. }\label{vertices}
\end{center}
\end{figure}

We can rewrite  (\ref{LDA}) and define the Dirac anisotropic Lagrangian
\begin{eqnarray}\label{lag-anisoD}
{\cal L_D}=\bar\psi (i \gamma^0\partial_0)\psi+b_1\bar\psi (i \gamma^i\partial_i)\psi+b_2\bar\psi (i \gamma^i\partial_i)^2\psi-m\bar\psi\psi,
\end{eqnarray}
 and, up to irrelevant surface terms, the interaction Lagrangian
\begin{eqnarray}\label{lag-anisoI}
{\cal L_I}&=&e\bar\psi(\gamma^0A_0+b_1\gamma^iA_i)\psi+e^2b_2\bar\psi(\gamma^iA_i)^2\psi+ieb_2(\bar\psi\gamma^j\gamma^i\partial_i\psi-\partial_i\bar\psi\gamma^i\gamma^j\psi)A_j,\nonumber\\
&=&V_1+V_2+V_5+V_4+V_3,
\end{eqnarray}
such that the total Lagrangian is ${\cal L_{T}}={\cal L_M}+{\cal L_{F}}+{\cal L_D}+{\cal
L_I}$. In (\ref{lag-anisoI}) we introduced a notation for the
vertices where, $V_1=e\bar\psi(\gamma^0A_0)\psi,{
V_2}=eb_1\bar\psi(\gamma^iA_i)\psi$, etc. These vertices, fixed by
${\cal L_{I}}$,  are graphically represented in
Fig.(\ref{vertices}).  For the free  fermion propagator we obtain
\[
S(k)=\frac{i\left(\hat{\not\!{k}}+b_1\bar{\not\!{k}}+b_2\textbf{k}^2+m\right)}{k_0^2-(b_1^2+2Mb_2)\textbf{k}^2-b_2^2\textbf{k}^4-m^2},
\]
where the hat  and the bar denotes the time and space components, respectively, i.e.,  $\hat{\not\!{k}}=\gamma^0k_0$, $\bar{\not\!{k}}=\gamma^ik_i$ and $\textbf{k}^2=k^ik^i$ (we will also use the notations $\hat{k}_0=k_0$ and $\bar{k}_i=k_i$, thus $\bar{k}^2=\bar{k}^i\bar{k}_i$).

\begin{figure}[h]
\begin{center}
\includegraphics[height=5cm]{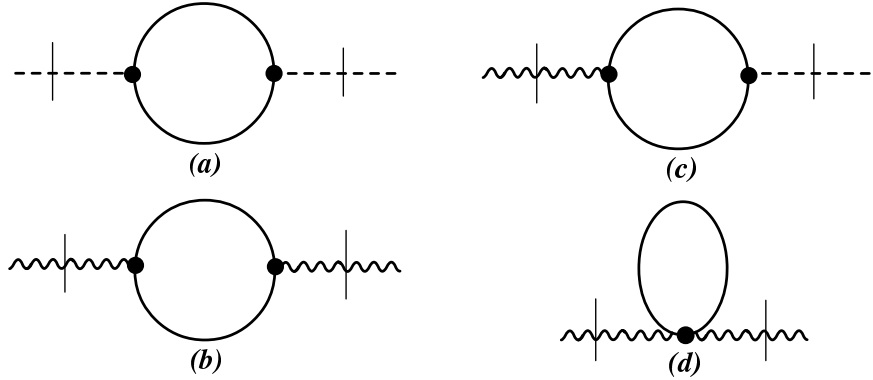}
\caption{One-loop contributions to the gauge field self-energy. The figures (a), (b) and (c) involve two triple vertices with external fields $(A_0,A_0)$,  $(A_0,A_i)$ and $(A_i,A_j)$, respectively. The diagram (d) contains  the quadrilinear vertex with $(A_i,A_j)$ external fields.}\label{2pts}
\end{center}
\end{figure}

The corrections to the gauge field propagator at one loop, are represented in Fig.(\ref{2pts}) whose analytical expressions are
\begin{eqnarray}
\Pi_{ab}^{\mu\nu}=  C_{ab}\int d^z\hat{k}d^d\bar{k}Tr[V_a^{\mu}S(k)V_b^{\nu}S(k+p)],\\
\Pi_{5}^{ij}=   C_5\int d^z\hat{k}d^d\bar{k}Tr[\gamma^i\gamma^jS(k)],
\end{eqnarray}
\noindent where $V_a^{\mu}$ is the time or spatial component $(\mu=0,i)$ associated to the vertices defined as in~(\ref{lag-anisoI}), thus $a,b=1,...,4$ and $C_{ab}$ is the momentum independent coefficient correspondent to each amplitude. The development of these expressions are discussed in appendices \ref{appendixB} and \ref{appendixA}.

By power counting we find: the diagram (\ref{2pts}\textit{c}) is linearly or logarithmically divergent if the vertex attached to the wavy line has or does not have a derivative; the divergences of the diagram  (\ref{2pts}\textit{b}) are quadratic if both vertices have one derivative, linear when one vertex has a derivative and none in the other and finally logarithmic when none derivative appears in the vertices just like in the Fig. (\ref{2pts}\textit{a}), which is composed by two external $A^0(p)$ components. 

In the sequel we will consider the small momenta regime and Taylor expand the self-energy corrections. The usual CS term (i.e. the first term in (\ref{CSexpec}))  is obtained from the first order terms on the Taylor derivative expansion while the extended CS term (which have a form like the second and third terms in (\ref{CSexpec}))  is given by the third order terms. Considering that the highest divergence is quadratic, the  extended CS contributions, which are of third order on Taylor expansion, are of course finite. Differently, for the usual CS terms, the algebra reduces the degree of divergence of some diagrams only  to logarithmic so that they requires the introduction of a regularization scheme. Similarly to the relativistic theory, in the sense that there is a regularization dependence, there is an ambiguity on the induced CS term.  

For the even terms in the momentum expansion, which do not contribute to the CS terms, the integration on momenta furnishes hypergeometric functions, as  indicated in the appendix \ref{appendixB}.  In this case, the divergences are characterized by the Schwinger parameter, $x$, which receives contributions from the hypergeometric and from the coefficients associated to them (see Eq. \ref{hypergeo}). To integrate on the $x$ parameter, we will power expand the hypergeometric for small $x$ and  $d=2$, until its exponent becomes nonnegative. Observe  that,  here the isotropic limit, $b_2\rightarrow 0$, cannot be  taken  singly because it is inconsistent with the small $x$ expansion. However, although the divergent terms individually exhibit poles for $d=2$,  they are canceled when summed to produce the total self-energy correction, allowing  for a smooth isotropic limit.


\section{Transversality of the gauge self-energy}\label{current}

The conservation of the  Noether's  current 
\begin{eqnarray}\label{cor00}
J^0=\bar\psi\gamma^0\psi,
\end{eqnarray}
\[
J^k= b_1\bar\psi\gamma^k\psi-ib_2\left[(\partial_i\bar\psi)\gamma^i\gamma^k\psi-\bar\psi\gamma^k\gamma^i(\partial_i\psi)\right]+eb_2\left[\bar\psi\gamma^i\gamma^k\psi+\bar\psi\gamma^k\gamma^i\psi\right]A_i\;.
\]
allows us to prove the transversality of the self-energy corrections. Indeed,
 we may write the ${\cal L}_{I}$ in terms of the  above  current
\begin{eqnarray}
{\cal L}_{I}=\frac{e}{2}\left[A_{\mu}J^{\mu}_{(A^{\rho}=0)}+A_{\mu}J^{\mu}\right],
\end{eqnarray}
where in the first term, the argument of the current $(A^\rho)$  must be taken equal to zero. 

Now, in computing of the self-energy contributions notice that it is equal to 
\begin{eqnarray}\label{CUcons}
\langle T\frac{\delta {\cal S}_{I}}{\delta A_{\mu}}\frac{\delta {\cal S}_{I}}{\delta A_{\nu}}\rangle\propto\langle J^{\mu} J^{\nu} \rangle\; ,
\end{eqnarray}
where ${\cal S}_{I}=\int{\cal L}_{I}$ is the interaction action; (\ref{CUcons}) must be transversal due to the conservation of the current.


\section{Concluding remarks}\label{remarks}

In this work we studied the effects of the anisotropy of the spacetime on the generation of the CS term and on the self-energy correction to the gauge field propagator.  We proved that this correction is transversal by constructing the conserved Noether's current which interacts with the gauge field. 

Some interesting aspects of the calculation are related to the structure of the divergences. For the even part of the gauge field self-energy, there are seven amplitudes involving the new vertices which are divergent  and their pole terms break gauge invariance. Nevertheless, a contribution coming from the usual vertices cancel these divergences. 

There are two types of CS terms that may be induced by the radiative corrections,  the usual CS term
which contains just one derivative and the extended CS term with three derivatives. Whereas the second type of CS term is always finite, the usual  one  presents results which are divergent  but these divergences notabily cancel among themselves when the total contribution is considered. Furthermore, the coefficient of the usual CS term  is regularization dependent and vanishes if dimensional reduction is adopted.  In this situation, one is forced to consider the extended CS term which constitute then the dominant contribution.

By referring to the usual CS terms, notice that the contributions coming from the usual vertices are finite, therefore free from ambiguities caused by the regularization. On the other hand, the contributions coming from the new vertices are regularization dependent. Besides that,  the explicit $b_2$ factor  from that vertices  is canceled because it is multiplied by amplitudes which diverge as $b_{2}\rightarrow 0$, this furnishes non zero contributions when the isotropic limit is considered. Consequently, the regularization dependence is restored and in the isotropic limit unexpected results, as the just mentioned vanishing of the usual CS term if dimensional reduction is employed, are obtained. 

Considering the extended CS, all amplitudes associated to it are finite even if $b_{2}\rightarrow 0$. The isotropic limit for these terms is smooth remaining only the contributions coming from the usual vertices.

By considering the Eqs. (\ref{geralij}) and (\ref{geral0i}), the induced CS terms can be written as
\begin{eqnarray}
\Pi^{\mu\nu}_{CS}=-\frac{\left[\alpha_{\mu\nu}+\hat{p}^2+\bar{p}^2(b_1^2+4mb_2)\right]e^2b_1^2\epsilon^{\mu\nu\rho}p_{\rho}}{48\pi m^2(b_1^2+4mb_2)}
\end{eqnarray}
in accord with our proposal (\ref{CSexpec}). Notice that the breaking of the Lorentz invariance is a very simple function of $b_1$ and $b_2$ and that in the isotropic limit with $b_1=1$ the Lorenz symmetry is restored. 
\acknowledgments
\vskip.5cm

\noindent This work was partially supported by Funda\c c\~ao de Amparo \`a Pesquisa do Estado de S\~ao Paulo (FAPESP), Conselho Nacional de
Desenvolvimento Cient\'ifico e Tecnol\'ogico (CNPq) and { Coordenac\~ao de Aperfei\c coamento de Pessoal de N\'ivel Superior (CAPES)}.  {(FP) acknowledge the support to this research by Direcci\'on de Investigaci\'on y Desarrollo de la Universidad de La Frontera  and Facultad de Ingenier\'ia, Ciencias y Administraci\'on, Universidad de La Frontera (Temuco-Chile) .}

\appendix

\section{The calculation procedure}\label{appendixB}

In this appendix we will describe the procedure employed to calculate the Feynman diagrams. As an example, we will consider the amplitude given by the vertices $(eb_1\bar\psi\gamma^i\psi A_i)$ and $(eb_1\bar\psi\gamma^j\psi A_j)$ which leads us to:
\begin{eqnarray}\label{exemp}
\Pi^{ij}(p)=\frac{(eb_1)^2}{(2\pi)^3}\int d\hat{k}d^d\bar{k}\;\textrm{Tr}[\gamma^iS(k)\gamma^jS(k-p)].
\end{eqnarray}
 To solve (\ref{exemp}) we adopt the dimensional reduction scheme, in which all the algebra of the gamma matrices are done in $d=2$ and afterwards the integral is promoted to $d$  dimensions~\cite{Siegel:1979wq}. Therefore, the leading term in the Taylor expansion, i.e. the term with $(\hat{p}=0,\bar{p}=0)$ is
\begin{eqnarray*}
-\frac{(eb_1)^2}{(2\pi)^3}\int d\hat{k}d^d\bar{k} \left[\frac{2g^{ij}[m^2-\hat{k}^2-b_2\textbf{k}^2(-\textbf{k}^2b_2-2M)]}{(\hat{k}^2-\textbf{k}^2(b_1^2+2Mb_2)-\textbf{k}^4b_2^2-m^2)^2}\right].
\end{eqnarray*}
To perform  the above integral, we use the Schwinger's representation and  integrate on the momenta. The result is very extensive, therefore we will consider just one of the terms which leads to a divergent result:
\begin{eqnarray}\label{hypergeo}
e^{-iM^2x}x^{-d/4} \phantom{.}_1F_1\left(\frac{d+2}{4};\frac{1}{2};\frac{ix(b_1^2+2Mb_2)^2}{4b_2^2}\right),
\end{eqnarray}
where $x$ is the Schwinger parameter and $_1F_1$ denotes the hypergeometric function. Note that this function is divergent in the limit $x\rightarrow 0$, so we will Taylor expand it, up to a nonnegative power of $x$ for $d=2$. Thus we obtain
\begin{eqnarray}
x^{-d/4} \phantom{.}_1F_1\left(\frac{d+2}{4};\frac{1}{2};\frac{ix(b_1^2+2Mb_2)^2}{4b_2^2}\right)\rightarrow x^{-d/4}\left(1+2\frac{d+2}{4}\frac{ix(b_1^2+2Mb_2)^2}{4b_2^2}\right)
\end{eqnarray}
The final result is obtained by integrating on $x$ and expanding the result around $d=2$, such that the general results for this amplitude is given by 
\begin{eqnarray}
\frac{ie^2b_1^2g^{ij}}{2\pi b_2(d-2)}+\textrm{finite}\label{pi22div}.
\end{eqnarray}
Notice that this divergence is absent in the relativistic theory because it clearly comes from modifications introduced by the anisotropy.  Observe also that  from the argument of the hypergeometric function in (\ref{hypergeo}) that the isotropic limit, $(b_2\rightarrow 0)$, is incompatible with the adopted expansion for $x\rightarrow 0$.

On the other hand, if we consider the subleading term in the Taylor expansion, responsible for the usual CS term, i.e. $\frac{\partial\Pi^{ij}}{\partial{\hat{p}}}\Big|_{\hat{p}=\bar{p}=0}$, the result is
\begin{eqnarray}\label{hyperfinite}
\frac{(eb_1)^2}{(2\pi)^3}\int d\hat{k}d^d\bar{k}\left[\frac{2i\epsilon^{ij0}(m-b_2\bar{k}^2)}{(\hat{k}^2-\textbf{k}^2(b_1^2+2Mb_2)-\textbf{k}^4b_2^2-m^2)^2}\right].
\end{eqnarray}
To compute the above integral, which is finite by power counting,  we simply take $d=2$ and integrate on the momenta. 

\section{Interaction vertices}\label{appendixA}

In this appendix we will present  the one loop contributions, for small momenta, to the self-energy of the gauge field. From the interaction Lagrangian we observe that we have a total of 12 different amplitudes coming from the Wick's contractions of the vertices. The sum of all contributions gives (further details will be presented elsewhere):   
\begin{eqnarray}\label{geralij}
\Pi^{ij}_{CS}= -\frac{\left[\alpha_{ij} + \bar{p}^2(b_1^2+4mb_2)+\hat{p}^2\right]e^2b_1^2\epsilon^{ij0}\hat{p}_0}{48\pi m^2(b_1^2+4mb_2)}
\end{eqnarray}
and
\begin{eqnarray}\label{geral0i}
\Pi^{0i}_{CS}=-\frac{\left[\alpha_{0i}+\bar{p}^2(b_1^2+4mb_2)+\hat{p}^2\right]e^2b_1^2\epsilon^{0ia}\bar{p}_a}{48\pi m^2(b_1^2+4mb_2)}
\end{eqnarray}
where $\alpha_{ij}$ and $\alpha_{0i}$ are constant parameters introduced in (\ref{geralij}) and (\ref{geral0i}), respectively, to denote the ambiguity coming from the regularization scheme. We may note that there is a cancellation of the usual CS term, remaining only regularization dependent terms. This occurs due to the contributions of the new vertices introduced by the anisotropy and by the fact that their $b_2$ vertex factor is eliminated after performing the momenta integrals.

In the isotropic limit $b_2\rightarrow 0$ of  (\ref{geralij}) and (\ref{geral0i}) all the extended CS terms coming from the new vertices contributions are cancelled and we get
\begin{eqnarray}
\Pi^{ij}_{CS}=-\frac{e^2\left(\alpha_{ij}+b_1^2\bar{p}^2+\hat{p}^2\right)\epsilon^{ij0}\hat{p}_0}{48\pi m^2}\nonumber
\end{eqnarray}
and
\begin{eqnarray}
\Pi^{0i}_{CS}=-\frac{e^2\left(\alpha_{0i}+b_1^2\bar{p}^2+\hat{p}^2\right)\epsilon^{0ia}\bar{p}_a}{48\pi m^2}.\nonumber
\end{eqnarray}
 By taken $b_1=1$ we obtain an expression similar to that one introduced in \cite{Deser:1999pa}.


\end{document}